
%
%
%

\def\cl{\centerline}
\def\over{\hskip 200pt}
\def\quotation{\narrower\noindent}
\parskip=5pt

\ \

\noindent {\it Comments on Astrophysics}: in press

\bigskip

\cl{\bf A Century of Gamma Ray Burst Models }

\bigskip
\baselineskip=9pt

\noindent More than 100 gamma-ray burst progenitor models have now been
published in refereed journals. A list of the models published before the
end of 1992 is presented and briefly discussed.  The consensus of the
present astronomical community remains that no specific model is
particularly favored for cosmic bursts.  Recent BATSE results make most of
these models untenable in their present form, opening up the field to
another era of speculative papers. Is speculation in this area becoming
valueless? Alternatively, one may argue that the new data makes many of the
old models untenable, and simply adapting old models may not be sufficient.
With this in mind, three relatively unexplored ``toy" paradigms are
suggested from which more detailed models for the progenitors of gamma-ray
bursts may be made.

\bigskip


\bigskip\bigskip
\bigskip\bigskip

{\bigskip\quotation\narrower\baselineskip=10pt
``For theorists who may wish to enter this broad and growing field, I
should point out that there are a considerable number of combinations, for
example, comets of antimatter falling onto white holes, not yet claimed."
\hfil\break\over
\ \ \ \ \ \ \ \ \ \ \ \ \ \ \ \ \ \ \ \  - M. Ruderman$^1$  }

\bigskip\bigskip
\baselineskip=10pt

\noindent 1. INTRODUCTION

\bigskip

\noindent Gamma-ray bursts (GRBs), discovered 25 years ago$^2$, remain
one of the biggest mysteries in modern astronomy.  No theoretical model
explaining GRBs has gained general acceptance, although now more than 100
have been proposed in the refereed literature.  Is speculation in this
field becoming valueless?  Is there any value to past speculative models
that no longer can be considered reasonable GRB models in light of present
data?

A list of the more than 100 papers suggesting GRB progenitor models, or
variants thereof, is given in Table 1.  A paper will appear on this list if
it was published in a refereed journal, appeared before the end of 1992,
and proposed a new or revised model for the origin of a GRB. A reasonable
effort was made to make Table 1 complete, however it is probable that
several papers were missed.  A conscience subjective decision {\it not} to
included a paper might have been made if it was deemed that the paper did
not create a significantly new progenitor model or did not add
significantly to an existing progenitor model (the paper might still be an
excellent scientific paper, however). A paper might also have been excluded
if it focussed specifically on the physics of a potential mechanism rather
than suggesting a significantly different type of mechanism. Papers
appearing in journals less circulated in the United States might also have
been missed. Please note that the numbers of references in Table 1 are
different from the reference numbers of papers cited in the reference
section at the end of this paper.

Table 1 is divided into 8 columns.  The first column gives a model reference
number.  Models are listed in chronological order of the date they were
received by the journals.  In the case of two models received on the same
day, the model that was published first is listed first.

Column two lists the lead author.  If there were two or more authors, an
``et al." is given following the first author. Column 3 lists the year the
article was published. Column 4 lists the reference in a compact form. Most
of these conform to the accepted modern format, however several
non-standard abbreviations were made, primarily due to space limitations.
Specifically,
``CJPhys" refers to the Canadian Journal of Physics,
``CosRes" refers to Cosmic Research,
``PRL" refers to Physical Review Letters,
and ``SovAstron" refers to Soviet Astronomy.

Column 5 lists the major progenitor body involved in the GRB model. Many
abbreviations are straightforward with ``NS" meaning neutron star, ``WD"
meaning white dwarf, ''BH" meaning black hole and ``AGN" meaning active
galactic nucleus.  Less standard abbreviations are:
``CS" meaning cosmic strings,
``DG" meaning dust grain,
``GAL" meaning external galaxy,
``MG" meaning magnetic reconnection,
``RE" meaning relativistic elections,
``SS" meaning strange star,
``ST" meaning normal star,
and ``WH" meaning white hole.

If a second body is involved in the GRB model, it is listed in column 6,
even if it cannot be clearly labelled as a body.
The following additional abbreviations were used:
``AGN" meaning active galactic nucleus,
``AST" meaning asteroid,
``COM" meaning comet,
``ISM" meaning interstellar medium,
``MBR" meaning microwave background,
``PLAN" meaning planet,
and ``SN" meaning supernova shock.

Column 7 lists the location of the GRB explosion. Here
``COS " refers to a cosmological setting,
``DISK" refers to the disk of our Milky Way Galaxy,
``HALO" refers to the halo of our Galaxy,
and ``SOL" refers to the outer solar system.  In cases where the GRB
location was not well specified between the Galactic disk and the Galactic
halo, the later location was typically chosen if energy constraints
allowed.

Column 8 gives a brief description of the model (or refinement) proposed. I
apologize for the gross generalizations made here and for any inaccuracies.
Several times terms and abbreviations are used in the description that need
explanation, and I must ask the reader to consult the papers cited for this
explanation. Happily, I was not shaken from my belief that once terms are
defined, the gist of any good scientific paper can be summarized in five
words or less. (Admittedly this makes a better parlor game than a truism.)

The first entry in Table 1 requires explanation.  The prime reason the GRB
discovery paper$^{2}$ gives for the initial search was to test the
prediction of GRB existence made by Colgate in Reference 1 of Table 1. GRBs
were discovered in this search but found {\it not} to be coincident with
observed supernovae in local galaxies, as this Colgate model predicts.
However, since this Colgate paper gave a model for GRBs which fostered GRB
detection, it is arguably the first GRB model, even though it predates
their detection.

Inspection of Table 1 shows several interesting trends. First of all, most
of the models are based in the Galactic disk, and most are based on neutron
stars. The most diverse group of models was published immediately after the
discovery of GRBs, but over the years a wide variety of distinctly
different models have been published.  Based on the publication record, it
appears that the community had generally settled on the idea of Galactic
disk neutron star progenitors in the 1980s, as the majority of papers
published then were refinements of this idea (with a few very notable
exceptions). Any settling that might of occurred then became unsettled with
the announcement of the first BATSE results in September 1991.

Launched in April of 1991, the {\it Compton} Gamma Ray Observatory
incorporated an instrument specifically designed to detect and measure
GRBs: the Burst and Transient Source Experiment (BATSE).  In September of
1991 at The Compton Observatory Science Workshop$^{3}$ in Annapolis
Maryland, the BATSE team, headed by Dr. Gerald Fishman, briefly summarized
the results of the first few months of BATSE observations. These results
shook gamma-ray burst understanding to the core.  They showed that BATSE
had so far measured an angularly isotropic GRB distribution, and that the
brightness distribution (sometimes called ``log N - log S") was not
uniformly a -3/2 power law.  Although this data was not in conflict with
previous data, most scientists and scientific modeling had predicted an
angular distribution in which the galactic plane was visible, and, barring
this, a continuation of the -3/2 power law results. These results were
first published in a paper sent to Nature.$^{4}$

The result of this announcement was a dramatic shift in GRB modeling.
Papers submitted in 1992 were, for the first time, predominantly
cosmologically placed.  There was also a slight shift away from neutron star
progenitor models, although NS models still outnumbered all others models
combined.

The GRB progenitor problem is now arguably the most prolific in
astronomical history, easily surpassing the pulsar problem in this
area.$^{5}$ (For a list of potential pulsar progenitor models, which
numbered 20 about 3 years after their discovery, see Table 2 of Ref. 5.
Even at this early date, though, there was a very strong community
sentiment toward rotating neutron stars.) Reasons for this include the
uncertainty of several important data features, the relatively long period
of speculation, the relatively large amount of data needed to solve the
dilemma compared to the amount of data taken, the relatively large numbers
of astronomers and astronomical journals in the world today, the relative
ease which word processing makes published speculation possible, and the
pressure to publish in today's academic environment, to name a few.
Probably the best reason for the proliferation of GRB models, though, is
that new data entering the field has not bolstered any specific model (or
even setting). The BATSE results, in fact, have made the majority of
published models more tenuous. Therefore, in light of this new data, it is
possible that even this list of over 100 models lacks diversity.

\bigskip\bigskip

\noindent 2. SERENDIPITOUS MODELS

\bigskip

\noindent Speculative model building based on reliable data {\it can} be
good science, and should be encouraged so long as it a) reasonably explains
the data, b) is falsifiable and/or c) is generally interesting
astrophysically in its own right and potentially applicable to areas
outside GRBs. Any model that does either a, b, or c particularly well
should be considered by itself meritorious.

There are models or types of speculative modeling that should not be so
encouraged.  Models that use non-standard physical laws, that incorporate
astronomical objects that are not known to exist, or that rely on data
that is not well understood should be treated with extra scrutiny. These
models should only be proposed if they are particularly falsifiable.
Otherwise, even if they are correct, few will believe them.

In every flurry of scientific speculation, inevitably most of this
speculation is wrong.  However, even if the speculation did not ultimately
result in a viable model for the mystery in question, many times the
speculation was valuable in its own right.  In a way, this is the
theoretical equivalent to serendipitous observational discovery.

Is there the potential for theoretical serendipity in speculation on the
origins of GRBs?  Hopefully there will be numerous examples in the GRB
model list. Here two potentially interesting cases are suggested.

The first is that of colliding neutron stars (see reference 89 in Table 1,
and all subsequent references with ``NS" listed in both columns 5 and 6).
These models may still turn out to be the correct model for GRBs, but even
if they are not, they could still be valuable as serendipitous speculation.
Neutron star binary systems are known to exist, and the stars are known to
be spiraling toward each other while releasing binding energy in the form
of gravitational waves.$^{6}$ Therefore neutron star - neutron star
collisions must happen occasionally - the questions are at what rate and at
what visibility.  Possibly these collisions would not be observable as
classical GRBs, but in other radiations, and with another frequency.

A second case involves models proposed to explain the majority of GRBs may
come in useful in understanding a smaller class of similar bursts: soft
gamma repeaters (SGRs).$^{7}$ Particularly exemplary in this regard might
be the models of accretion and thermonuclear detonation on the surface of
neutron stars (see references 7 and 27 of Table 1, and many subsequent
papers).  These models generally do not release enough energy to account
for cosmic GRBs at cosmological distances, but may release enough energy to
explain SGRs.

\bigskip\bigskip

\noindent 3. A BIASED GRB MODEL START UP KIT

\bigskip

\noindent
Why the continued emphasis on neutron star models, particularly in the
light of the new data?  For one reason, neutron stars still represent
active environments where energy fields go to extremes, creating a ripe
setting for the powerful explosion that are GRBs. Also, it is a tempting
coincidence that if GRBs lie at cosmological distances, the energy released
in a GRB (assuming isotropy of the explosion), is a few percent of the
binding energy of a neutron star.

One general problem with neutron star models is that they generally
liberate most of the energy in neutrinos. To complement such a model, a
method of converting a fraction (even 0.1 percent) of the neutrinos into
$\gamma$-rays must also be found.  Several of the most recent models have
concerned themselves with this (for example reference 115 of Table 1).

No single piece of evidence has been found to suggest neutron stars
conclusively as GRB progenitors. Were a rotation period reminiscent of a
pulsar found in a GRB light curve, this could be considered conclusive
evidence. Therefore, to reflect this lack of evidence, two of the paradigms
suggested below are not constrained to neutron star environs.

In generating a model of any type based upon believable data, one must know
which data to believe.  This is particularly difficult in GRB astronomy, as
there is continuing debate as to whether the GRB data show cyclotron lines,
annihilation lines, a distribution which is truly isotropic, a duration
histogram indicative of one population or two, whether SGRs indicate a
separate class of should be grouped with the other GRBs, or whether optical
counterparts have ever been seen. As noted above, this diversity may be
partially responsible for the large number of GRB models, as a different
model is usually needed to explain a different subset of the data.

Deciding which data to believe is implicitly a biased procedure. Therefore,
before stating any new idea on origins of GRBs, I state my prejudices
explicitly below.  Let me start by stating here that I find my biases
change in an unscientific manner, depending on how much data I perceive
supports a particular bias, the quality and manner this data was taken, the
history of a data set, the general biases in the literature, the specific
biases of my closer colleagues, and to whom I have listened most recently.
The biases listed below are indicative of no one other than myself.  Some
are more controversial than others.

\noindent
$\bullet$ SGRs are a different class of GRB and are not to be explained by
the cosmic GRB model.  (Note that models trying specifically to explain
SGRs {\it are} included in Table 1.)

\noindent
$\bullet$ A model must predict an isotropic but ``confined" (Log N vs. Log
S not fully described by a -3/2 power law) GRB distribution that is
consistent with the current BATSE data.

\noindent
$\bullet$ A model must predict that each GRB has a somewhat similar
spectrum.  Generally this means increasing in the hard X-ray, turning
over in the early gamma-ray, decreasing power law in the gamma-ray, and
turning off in the hard gamma-ray.

\noindent
$\bullet$ A model geometry must be able to explain the choppy time
structure inherent in the data. Therefore I preferred that something during
the process should either come in pieces or break up into pieces.  This is
because the time profiles of GRBs can be quite complicated and composed of
many discernable sub-pulses.$^{8}$

\noindent
$\bullet$ The GRB process should occur at cosmological distances, and
additionally, should be uniformly distributed in the universe.  This is
because the log N - log S relationship and time-dilation GRB comparisons
fits a uniform cosmological distribution quite well.$^{9-11}$

\noindent
$\bullet$ Neutron star models are to be avoided. Most of the literature is
composed of neutron star models, and most probably if GRBs are formed in
neutron star environs, one of the existing models already goes most of the
way to explaining it.  Besides, no convincing periodicities indicative of
neutron star rotation have ever been found.

\noindent
$\bullet$ No antimatter.  I feel that there is no strong evidence that a
substantial part of the universe is composed of antimatter.

\noindent
$\bullet$ Relativistic beaming should be a natural consequence of the
model.$^{12}$  This is to stop $\gamma-\gamma$ interactions from degrading
the higher energy tail of GRB's spectrum.

\noindent
$\bullet$ Models should be capable of producing time scales as short as a
millisecond and as long as 5 minutes.  These roughly correspond to the
duration of the longest and shortest GRBs.

\noindent
$\bullet$ Models should not predict GRB recurrence in the same angular
location. GRBs do not recur at the same place in the sky, at least not on
the time scale from 10 minutes to a few years.  Note however that GRBs do
show recurrent pulses on the time scale of a few 100s of seconds.

\bigskip\bigskip

\noindent 4. THREE MORE PARADIGMS

\bigskip

\noindent
One might think that with 118 GRB progenitor models in the journals, no
more are needed.  One may also argue that the new data makes many of the
old models untenable, and simply adapting old models may not be sufficient.
There are many extremely energetic places and phenomena already known in
astrophysics and surely many more yet still to be discovered.  In this
light, it appears that the current GRB model list may not be diverse
enough.  Many similar models may not be as useful as a few very different
ones.  Is it possible to build completely different models and paradigms
that fit the data and yet still rely on plausible astronomical settings and
established physical laws?

The progenitor paradigms that follow are not well detailed: they
are at most toys or outlines from which more elaborate models can be
built.  I don't fully believe any of them, but there are aspects to each
of them I find appealing. They are provided as examples and as ``food for
thought."  My hope is that they will at least foster discussion and more
diverse model building.

\noindent {\bf Lightning: }
Although there can be many models based on lightning, here is one that
tries to be cosmological. Lightning occurs frequently in planetary
settings, causing one to wonder how frequently it occurs outside such a
setting. Lightning has recently been suggested in a more general
astrophysical context$^{13}$.

There is known to be at least a little bit of intergalactic matter in the
form of un-ionized baryonic matter, some of which is known to be clumped
into higher density clouds.$^{14}$  General gravitational settling combined
with collisions of material in these clouds could lead to non-negligible
E-fields and charge separation.  As two sub-clumps pass near each other, a
series of lightning bolts could discharge between the two. But here, unlike
on earth, the distance and voltage drop between the clouds would accelerate
charged particles to energies where they would beam radiation in the
gamma-ray band when they strike the destination cloud(s).

The good points of this paradigm include that it explains naturally the
lack of visible objects at GRB locations. The time series of the intensity
of terrestial lightning has similar properties to the time series of
intensity of GRBs.

Bad points of this paradigm include that a detailed theory would have to
rely on densities of clouds that are not well constrained by measurements,
sizes that are not known to several orders of magnitude, and gas and dust
properties that are completely unknown. Energy constraints are also too
ad-hoc: why wouldn't one get more energetic or less energetic lightning
bolts?  One must also fine-tune the initial conditions so that there is not
too much ionized matter around to damp the charge separation necessary for
lightning to be created.

Were such a paradigm correct, there will never be a definitive correlation
found between any bright object and a GRB location, no matter the angular
precision of the GRB location. There might, however, be a correlation
between the magnitude of absorption of QSO light and GRB locations, when
GRB locations are known to a few arcseconds or better. A similar but
smaller scale lightning effect should occur in stellar neighborhood
molecular clouds.

\noindent {\bf Deflection of AGN Jets: }
Brainerd (reference 114 of Table 1) has remarked on the similarity of
published models of AGN and the models needed for GRBs. One way to
facilitate this is for a comet (for example) to wander into an AGN jet and
scatter some of the beam temporarily toward us.  Soon the comet melts.
Comets are particularly good as deflectors since they may be composed of
several pieces, which could give rise to the pulse - composed structure of
GRBs.

On the positive side, many of the physical aspects of AGN models that GRB
models have in common are naturally explained.  The pulse-structure of GRB
time series may also be naturally explained by the piece-meal structure of
comets. However, one might expect that the total energy deflected by the
comets would be widely variable.  One might also expect that AGN deflection
models would have repeating GRBs, as different comets wander into the same
AGN jet. Brighter, longer AGN jets are more likely to cause GRBs, but these
are typically more distant, so GRBs would not be uniformly distributed.

Testable predictions of such a model include that when GRBs are better
located to an accuracy of better than 10 arcseconds, they should be
correlated with AGN positions.  Also, GRBs will be seen to occasionally
repeat from the same location, but with different light curves.

\noindent {\bf Mini-Black Holes Devouring Neutron Stars: } OK, I know I
said we don't need more neutron star models but this one was just too much
fun. This model was motivated by the ever-so-slim possibility that three
major astronomical puzzles could be solved in one fell swoop. Mini-black
holes (mBHs), on the order of a fraction of a solar mass, cannot be
excluded, presently, from comprising all the dark matter.  A few mBHs could
have found their way to the center of the sun to solve the solar neutrino
problem$^{15-16}$. What if one of these mBHs were to fall not into our Sun
but into a neutron star instead$^{17}$? One might expect a GRB.  As the mBH
ate its way through the star the mBH would increase in mass and eat faster.
After a while the neutron star would undergo massive restructuring (core
quakes) every pass of the mBH through it.  The tides of the mBH could also
cause explosive decompression on nearer parts of the neutron star.  The
neutron star could be ``eaten" completely on the time scale of a few
crossing times of the mBH, or just ``bitten" during one mBH unbound pass.
Neutron star - mBH collisions are unlikely to occur for uniformly
distributed chance encounters in the universe, but could be random
collisions in a dense stellar environment (near an AGN, for example), or
part of a binary system in a normal galaxy.

On the positive side is the extremely appealing idea that three major
astronomical problems could be solved simultaneously.  Also, in a gross
sense, the energy and timing considerations of this paradigm are roughly
OK.

However, the origin of mBHs is unclear at best.  Their present existence is
unsubstantiated and extremely ah-hoc.  One might expect neutron star
vibrations or rotation-induced periodicities to be evident in the GRB time
series, but they aren't. For lower mass mBHs, the mBH would have to take
many passes through the center of the neutron star before it got enough
mass to destroy the neutron star.  This oscillation time scale should be
evident in the GRB, and it isn't. Most of the neutron star binding energy
should be liberated in the form of neutrinos - one must still find a way to
convert a sizeable portion of the energy to gamma-rays.

Predictions of this paradigm include that fact that strong GRBs should have
detectable gravitational wave emissions in the next generation
gravitational wave detectors.  Also future arcsecond locations of strong
GRBs should show correlation with dim galaxies (which might house the
neutron stars). General mBH existence should be implied by the neutrino
emission of the Sun and the dynamics of some nearby stellar systems.

\bigskip

\noindent 5. DISCUSSION

\bigskip

\noindent  I apologize if I have omitted or badly described any models in
Table 1.  I will try to update Table 1 on a yearly basis, however, and
continually honor requests for a photocopy of it. Therefore, I welcome any
comments, corrections, or omissions that the reader may have on this table.

It will take more than speculation to solve the current GRB model
dilemma - it will certainly take more observations.  Clearly, several
observational uncertainties need to be resolved for theorists to know which
data subsets to believe. Do GRB show cyclotron lines, annihilation lines,
or repetition?  These questions should be answered by the current {\it
Compton} mission. Do GRBs show extra X-ray absorption in the galactic
plane?  Are GRB positions, when known more accurately, correlated with any
known object?  These are examples of questions which may be answered with
the next generation of GRB measuring instruments. If, when these data
arrive, they don't bolster an existing model, we may well be in for yet
another era of GRB model speculation!

\bigskip

\noindent {\bf Acknowledgements }

\noindent I thank Brad Schaefer for providing a preliminary list on which
Table 1 is based, and for many helpful comments and criticisms.  I also
thank the following people for humoring me through some unusual
discussions: Jay Norris, Thulsi Wickramasinghe, and Jerry Bonnell.

\bigskip
\parindent=0pt

{\bf References }

1. Ruderman, M., Ann. NY Acad. Sci. {\bf 262}, 164 (1975).

2. Klebesadel, R., Strong, I. B., Olson, R. A., Ap. J. 182, L85 (1973).

\hangindent=10pt
3. Shrader, C. R., Gehrels, N. and Dennis, B., {\it The Compton Observatory
Science Workshop}, (NASA conference publication 3137, 1992).

\hangindent=10pt
4. Meegan, C. A., Fishman, G. J., Wilson, R. B., Paciesas, W. S., Brock, M.
N., Horack, J. M., Pendleton, G. N., \& Kouveliotou, C., Nature {\bf 355},
143 (1992).

5. Hewish, A., Ann. Rev. Astron. Astrophysics {\bf 265}, (1970).

6. Taylor, J. H. \& Weinberg, J. M., Ap. J., {\bf 253}, 908 (1982).

\hangindent=10pt
7. Norris, J. P. et al., Ap. J. {\bf 366}, 240 (1991).

\hangindent=10pt
8. Higdon, J. C., \&\ Lingenfelter, R.E., Ann. Rev. Astron. Astrophysics
{\bf 28}, 401 (1990).

9. See Piran, T., Ap. J. {\bf 389}, L45 (1992);

10. Wickramasinghe W. A. D. T. et al., Ap. J. (submitted).

11. Fenimore E. E. et al., Nature (preprint).

\hangindent=10pt
12. Krolik, J. Pier, E. A., {\it Gamma-ray bursts - Observations, analyses
and theories}, (Cambridge, Cambridge University Press, 1992), 99.

\hangindent=10pt
13. Pilipp, W., Hartquist, T. W., \& Morfill, G. E., Ap. J. {\bf 387}, 364
(1992).

\hangindent=10pt
14. Sargent, W. L. W., Young, P. J., Boksenberg, A., and Tytler, D.,
Ap. J. Supp. {\bf 42}, 41 (1980).

15. Hawking, S. W., M. N. R. A. S. {\bf 152}, 75 (1971);

16. Clayton D. et al., Ap. J. {\bf 201}, 489 (1975).

17. Trofimenko, A. P., Ap. Space Sci. {\bf 168}, 277 (1990).

\bigskip
\parskip=0pt

\over ROBERT J. NEMIROFF

\over {\it NASA/GSFC/USRA}

\over {\it Greenbelt, MD 20771}

\vfill\eject

\font\sixrm=cmr6
\hsize=6.5 true in
\parskip=0pt
\parindent=0pt
\baselineskip=8pt
\nopagenumbers

\let\knuthitem=\item
\newcount\listnumber
\def\item{\global\advance\listnumber by 1 \knuthitem{\the\listnumber.}}

\bigskip

\centerline{ Table 1}

\bigskip

{\sixrm

\halign{ #\hfil& \quad#\hfil& \quad#\hfil& \quad#\hfil& \quad#\hfil&
\quad#\hfil& \quad#\hfil& \quad#\hfil \cr
\noalign{\smallskip\hrule\smallskip}
\noalign{\smallskip\hrule\smallskip}
\# \ \ & Author & Year & Reference &  Main & 2nd   & Place & Description \cr
       &        & Pub  &           &  Body & Body  &       &             \cr
\noalign{\smallskip\hrule\smallskip}
\cr
\item & Colgate            & 1968 & CJPhys, 46, S476      & ST &     & COS  &
SN shocks stellar surface in distant galaxy                              \cr
\item & Colgate            & 1974 & ApJ, 187, 333         & ST &     & COS  &
Type II SN shock brem, inv Comp scat at stellar surface  \cr
\item & Stecker et al.     & 1973 & Nature, 245, PS70     & ST &     & DISK &
Stellar superflare from nearby star                         \cr
\item & Stecker et al.     & 1973 & Nature, 245, PS70     & WD &     & DISK &
Superflare from nearby WD                                   \cr
\item & Harwit et al.      & 1973 & ApJ, 186, L37         & NS & COM & DISK &
Relic comet perturbed to collide with old galactic NS                 \cr
\item & Lamb et al.        & 1973 & Nature, 246, PS52     & WD & ST  & DISK &
Accretion onto WD from flare in companion             \cr
\item & Lamb et al.        & 1973 & Nature, 246, PS52     & NS & ST  & DISK &
Accretion onto NS from flare in companion             \cr
\item & Lamb et al.        & 1973 & Nature, 246, PS52     & BH & ST  & DISK &
Accretion onto BH from flare in companion             \cr
\item & Zwicky             & 1974 & Ap \& SS, 28, 111     & NS &     & HALO &
NS chunk contained by external pressure escapes, explodes         \cr
\item & Grindlay et al.    & 1974 & ApJ, 187, L93         & DG &     & SOL  &
Relativistic iron dust grain up-scatters solar radiation              \cr
\item & Brecher et al.     & 1974 & ApJ, 187, L97         & ST &     & DISK &
Directed stellar flare on nearby star                               \cr
\item & Schlovskii         & 1974 & SovAstron, 18, 390    & WD & COM & DISK &
Comet from system's cloud strikes WD                            \cr
\item & Schlovskii         & 1974 & SovAstron, 18, 390    & NS & COM & DISK &
Comet from system's cloud strikes NS                                  \cr
\item & Bisnovatyi- et al. & 1975 & Ap \& SS, 35, 23      & ST &     & COS  &
Absorption of neutrino emission from SN in stellar envelope           \cr
\item & Bisnovatyi- et al. & 1975 & Ap \& SS, 35, 23      & ST & SN  & COS  &
Thermal emission when small star heated by SN shock wave              \cr
\item & Bisnovatyi- et al. & 1975 & Ap \& SS, 35, 23      & NS &     & COS  &
Ejected matter from NS explodes                                       \cr
\item & Pacini et al.      & 1974 & Nature, 251, 399      & NS &     & DISK &
NS crustal starquake glitch; should time coincide with GRB         \cr
\item & Narlikar et al.    & 1974 & Nature, 251, 590      & WH &     & COS  &
White hole emits spectrum that softens with time                      \cr
\item & Tsygan             & 1975 & A\&A, 44, 21          & NS &     & HALO &
NS corequake excites vibrations, changing E \& B fields               \cr
\item & Chanmugam          & 1974 & ApJ, 193, L75         & WD &     & DISK &
Convection inside WD with high B field produces flare                 \cr
\item & Prilutski et al.   & 1975 & Ap \& SS, 34, 395     & AGN& ST  & COS  &
Collapse of supermassive body in nucleus of active galaxy     \cr
\item & Narlikar et al.    & 1975 & Ap \& SS, 35, 321     & WH &     & COS  &
WH excites synchrotron emission, inverse Compton scattering           \cr
\item & Piran et al.       & 1975 & Nature, 256, 112      & BH &     & DISK &
Inv Comp scat deep in ergosphere of fast rotating, accreting BH       \cr
\item & Fabian et al.      & 1976 & Ap \& SS, 42, 77      & NS &     & DISK &
NS crustquake shocks NS surface                                       \cr
\item & Chanmugam          & 1976 & Ap \& SS, 42, 83      & WD &     & DISK &
Magnetic WD suffers MHD instabilities, flares                         \cr
\item & Mullan             & 1976 & ApJ, 208, 199         & WD &     & DISK &
Thermal radiation from flare near magnetic WD                        \cr
\item & Woosley et al.     & 1976 & Nature, 263, 101      & NS &     & DISK &
Carbon detonation from accreted matter onto NS                        \cr
\item & Lamb et al.        & 1977 & ApJ, 217, 197         & NS &     & DISK &
Mag grating of accret disk around NS causes sudden accretion  \cr
\item & Piran et al.       & 1977 & ApJ, 214, 268         & BH &     & DISK &
Instability in accretion onto rapidly rotating BH                     \cr
\item & Dasgupta           & 1979 & Ap \& SS, 63, 517     & DG &     & SOL  &
Charged intergal rel dust grain enters sol sys, breaks up             \cr
\item & Tsygan             & 1980 & A\&A, 87, 224         & WD &     & DISK &
WD surface nuclear burst causes chromospheric flares                   \cr
\item & Tsygan             & 1980 & A\&A, 87, 224         & NS &     & DISK &
NS surface nuclear burst causes chromospheric flares                   \cr
\item & Ramaty et al.      & 1981 & Ap \& SS, 75, 193     & NS &     & DISK &
NS vibrations heat atm to pair produce, annihilate, synch cool          \cr
\item & Newman et al.      & 1980 & ApJ, 242, 319         & NS & AST & DISK &
Asteroid from interstellar medium hits NS                             \cr
\item & Ramaty et al.      & 1980 & Nature, 287, 122      & NS &     & HALO &
NS core quake caused by phase transition, vibrations                  \cr
\item & Howard et al.      & 1981 & ApJ, 249, 302         & NS & AST & DISK &
Asteroid hits NS, B-field confines mass, creates high temp            \cr
\item & Mitrofanov et al.  & 1981 & Ap \& SS, 77, 469     & NS &     & DISK &
Helium flash cooled by MHD waves in NS outer layers                   \cr
\item & Colgate et al.     & 1981 & ApJ, 248, 771         & NS & AST & DISK &
Asteroid hits NS, tidally disrupts, heated, expelled along B lines    \cr
\item & van Buren          & 1981 & ApJ, 249, 297         & NS & AST & DISK &
Asteroid enters NS B field, dragged to surface collision              \cr
\item & Kuznetsov          & 1982 & CosRes, 20, 72        & MG &     & SOL  &
Magnetic reconnection at heliopause                                   \cr
\item & Katz               & 1982 & ApJ, 260, 371         & NS &     & DISK &
NS flares from pair plasma confined in NS magnetosphere              \cr
\item & Woosley et al.     & 1982 & ApJ, 258, 716         & NS &     & DISK &
Magnetic reconnection after NS surface He flash                       \cr
\item & Fryxell et al.     & 1982 & ApJ, 258, 733         & NS &     & DISK &
He fusion runaway on NS B-pole helium lake                            \cr
\item & Hameury et al.     & 1982 & A\&A, 111, 242        & NS &     & DISK &
e- capture triggers H flash triggers He flash on NS surface        \cr
\item & Mitrofanov et al   & 1982 & MNRAS, 200, 1033      & NS &     & DISK &
B induced cyclo res in rad absorp giving rel e-s, inv C scat       \cr
\item & Fenimore et al.    & 1982 & Nature, 297, 665      & NS &     & DISK &
BB X-rays inv Comp scat by hotter overlying plasma                    \cr
\item & Lipunov et al.     & 1982 & Ap \& SS, 85, 459     & NS & ISM & DISK &
ISM matter accum at NS magnetopause then suddenly accretes            \cr
\item & Baan               & 1982 & ApJ, 261, L71         & WD &     & HALO &
Nonexplosive collapse of WD into rotating, cooling NS                 \cr
\item & Ventura et al.     & 1983 & Nature, 301, 491      & NS & ST  & DISK &
NS accretion from low mass binary companion                           \cr
\item & Bisnovatyi- et al. & 1983 & Ap \& SS, 89, 447     & NS &     & DISK &
Neutron rich elements to NS surface with quake, undergo fission   \cr
\item & Bisnovatyi- et al. & 1984 & SovAstron, 28, 62     & NS &     & DISK &
Thermonuclear explosion beneath NS surface                            \cr
\item & Ellison et al.     & 1983 & A\&A, 128, 102        & NS &     & HALO &
NS corequake + uneven heating yield SGR pulsations                    \cr
\item & Hameury et al.     & 1983 & A\&A, 128, 369        & NS &     & DISK &
B field contains matter on NS cap allowing fusion                     \cr
\item & Bonazzola et al.   & 1984 & A\&A, 136, 89         & NS &     & DISK &
NS surface nuc explosion causes small scale B reconnection        \cr
\item & Michel             & 1985 & ApJ, 290, 721         & NS &     & DISK &
Remnant disk ionization instability causes sudden accretion      \cr
\item & Liang              & 1984 & ApJ, 283, L21         & NS &     & DISK &
Resonant EM absorp during magnetic flare gives hot sync e-s   \cr
\item & Liang et al.       & 1984 & Nature, 310, 121      & NS &     & DISK &
NS magnetic fields get twisted, recombine, create flare               \cr
\item & Mitrofanov         & 1984 & Ap \& SS, 105, 245    & NS &     & DISK &
NS magnetosphere excited by starquake                                 \cr
\item & Epstein            & 1985 & ApJ, 291, 822         & NS &     & DISK &
Accretion instability between NS and disk                             \cr
\item & Schlovskii et al.  & 1985 & MNRAS, 212, 545       & NS &     & HALO &
Old NS in Galactic halo undergoes starquake                           \cr
\item & Tsygan             & 1984 & Ap \& SS, 106, 199    & NS &     & DISK &
Weak B field NS spherically accretes, Comptonizes X-rays       \cr
\item & Usov               & 1984 & Ap \& SS, 107, 191    & NS &     & DISK &
NS flares result of magnetic convective-oscillation instability       \cr
\item & Hameury et al.     & 1985 & ApJ, 293, 56          & NS &     & DISK &
High Landau e-s beamed along B lines in cold atm of NS            \cr
\item & Rappaport et al.   & 1985 & Nature, 314, 242      & NS &     & DISK &
NS + low mass stellar companion gives GRB + optical flash             \cr
\item & Tremaine et al.    & 1986 & ApJ, 301, 155         & NS & COM & DISK &
NS tides disrupt comet, debris hits NS next pass                      \cr
\item & Muslimov et al.    & 1986 & Ap \& SS, 120, 27     & NS &     & HALO &
Radially oscillating NS                                               \cr
\item & Sturrock           & 1986 & Nature, 321, 47       & NS &     & DISK &
Flare in the magnetosphere of NS accelerates e-s along B-field    \cr
\item & Paczynski          & 1986 & ApJ, 308, L43         & NS &     & COS  &
Cosmo GRBs: rel e- e+ opt thk plasma outflow indicated         \cr
\item & Bisnovatyi- et al  & 1986 & SovAstron, 30, 582    & NS &     & DISK &
Chain fission of superheavy nuclei below NS surface during SN         \cr
\item & Alcock et al.      & 1986 & PRL, 57, 2088         & SS & SS  & DISK &
SN ejects strange mat lump craters rotating SS companion              \cr
\item & Vahia et al.       & 1988 & A\&A, 207, 55         & ST &     & DISK &
Magnetically active stellar system gives stellar flare                \cr
\item & Babul et al.       & 1987 & ApJ, 316, L49         & CS &     & COS  &
GRB result of energy released from cusp of cosmic string           \cr
\item & Livio et al.       & 1987 & Nature, 327, 398      & NS & COM & DISK &
Oort cloud around NS can explain soft gamma-repeaters                 \cr
\item & McBreen et al.     & 1988 & Nature, 332, 234      & GAL& AGN & COS  &
G-wave bkgrd makes BL Lac wiggle across galaxy lens caustic     \cr
\item & Curtis             & 1988 & ApJ, 327, L81         & WD &     & COS  &
WD collapses, burns to form new class of stable particles             \cr
\item & Melia              & 1988 & ApJ, 335, 965         & NS &     & DISK &
Be/X-ray binary sys evolves to NS accretion GRB with recurrence    \cr
\item & Ruderman et al.    & 1988 & ApJ, 335, 306         & NS &     & DISK &
e+ e- cascades by aligned pulsar outer-mag-sphere reignition    \cr
\item & Paczynski          & 1988 & ApJ, 335, 525         & CS &     & COS  &
Energy released from cusp of cosmic string (revised) \cr
\item & Murikami et al.    & 1988 & Nature, 335, 234      & NS &     & DISK &
Absorption features suggest separate colder region near NS            \cr
\item & Melia              & 1988 & Nature, 336, 658      & NS &     & DISK &
NS + accretion disk reflection explains GRB spectra                   \cr
\item & Blaes et al.       & 1989 & ApJ, 343, 839         & NS &     & DISK &
NS seismic waves couple to magnetospheric Alfen waves                 \cr
\item & Trofimenko et al.  & 1989 & Ap \& SS, 152, 105    & WH &     & COS  &
Kerr-Newman white holes                                               \cr
\item & Sturrock et al.    & 1989 & ApJ, 346, 950         & NS &     & DISK &
NS E-field accelerates electrons which then pair cascade       \cr
\item & Fenimore et al.    & 1988 & ApJ, 335, L71         & NS &     & DISK &
Narrow absorption features indicate small cold area on NS             \cr
\item & Rodrigues          & 1989 & AJ, 98, 2280          & WD & WD  & DISK &
Binary member loses part of crust, through L1, hits primary           \cr
\item & Pineault et al.    & 1989 & ApJ, 347, 1141        & NS & COM & DISK &
Fast NS wanders though Oort clouds, fast WD bursts only optical       \cr
\item & Melia et al.       & 1989 & ApJ, 346, 378         & NS &     & DISK &
Episodic electrostatic accel and Comp scat from rot high-B NS        \cr
\item & Trofimenko         & 1989 & Ap \& SS, 159, 301    & WH &     & COS  &
Different types of white, ``grey" holes can emit GRBs                 \cr
\item & Eichler et al.     & 1989 & Nature, 340, 126      & NS & NS  & COS  &
NS - NS binary members collide, coalesce                              \cr
\item & Wang et al.        & 1989 & PRL, 63, 1550         & NS &     & DISK &
Cyclo res \& Raman scat fits 20, 40 keV dips, magnetized NS           \cr
\item & Alexander et al.   & 1989 & ApJ, 344, L1          & NS &     & DISK &
QED mag resonant opacity in NS atmosphere                             \cr
\item & Melia              & 1990 & ApJ, 351, 601         & NS &     & DISK &
NS magnetospheric plasma oscillations                                 \cr
\item & Ho et al.          & 1990 & ApJ, 348, L25         & NS &     & DISK &
Beaming of radiation necessary from magnetized neutron stars          \cr
\item & Mitrofanov et al.  & 1990 & Ap \& SS, 165, 137    & NS & COM & DISK &
Interstellar comets pass through dead pulsar's magnetosphere          \cr
\item & Dermer             & 1990 & ApJ, 360, 197         & NS &     & DISK &
Compton scattering in strong NS magnetic field                        \cr
\item & Blaes et al.       & 1990 & ApJ, 363, 612         & NS & ISM & DISK &
Old NS accretes from ISM, surface goes nuclear                        \cr
\item & Paczynski          & 1990 & ApJ, 363, 218         & NS & NS  & COS  &
NS-NS collision causes neutrino collisions, drives super-Ed wind  \cr
\item & Zdziarski et al.   & 1991 & ApJ, 366, 343         & RE & MBR & COS  &
Scattering of microwave background photons by rel e-s             \cr
\item & Pineault           & 1990 & Nature, 345, 233      & NS & COM & DISK &
Young NS drifts through its own Oort cloud                            \cr
\item & Trofimenko et al.  & 1991 & Ap \& SS, 178, 217    & WH &     & HALO &
White hole supernova gave simultaneous burst of g-waves from 1987A     \cr
\item & Melia et al.       & 1991 & ApJ, 373, 198         & NS &     & DISK &
NS B-field undergoes resistive tearing, accelerates plasma     \cr
\item & Holcomb et al.     & 1991 & ApJ, 378, 682         & NS &     & DISK &
Alfen waves in non-uniform NS atmosphere accelerate particles         \cr
\item & Haensel et al.     & 1991 & ApJ, 375, 209         & SS & SS  & COS  &
Strange stars emit binding energy in grav rad and collide           \cr
\item & Blaes et al.       & 1991 & ApJ, 381, 210         & NS & ISM & DISK &
Slow interstellar accretion onto NS, e- capture starquakes result  \cr
\item & Frank et al.       & 1992 & ApJ, 385, L45         & NS &     & DISK &
Low mass X-ray binary evolve into GRB sites                           \cr
\item & Woosley et al.     & 1992 & ApJ, 391, 228         & NS &     & HALO &
Accreting WD collapsed to NS                                          \cr
\item & Dar et al.         & 1992 & ApJ, 388, 164         & WD &     & COS  &
WD accretes to form naked NS, GRB, cosmic rays                       \cr
\item & Hanami             & 1992 & ApJ, 389, L71         & NS & PLAN& COS  &
NS - planet magnetospheric interaction unstable                       \cr
\item & Meszaros et al.    & 1992 & ApJ, 397, 570         & NS & NS  & COS  &
NS - NS collision produces anisotropic fireball                       \cr
\item & Carter             & 1992 & ApJ, 391, L67         & BH & ST  & COS  &
Normal stars tidally disrupted by galactic nucleus BH                 \cr
\item & Usov               & 1992 & Nature, 357, 472      & NS &     & COS  &
WD collapses to form NS, B-field brakes NS rotation instantly   \cr
\item & Narayan et al.     & 1992 & ApJ, 395, L83         & NS & NS  & COS  &
NS - NS merger gives optically thick fireball              \cr
\item & Narayan et al.     & 1992 & ApJ, 395, L83         & BH & NS  & COS  &
BH - NS merger gives optically thick fireball              \cr
\item & Brainerd           & 1992 & ApJ, 394, L33         & AGN& JET & COS  &
Synchrotron emission from AGN jets                                    \cr
\item & Meszaros et al.    & 1992 & MNRAS, 257, 29P       & BH & NS  & COS  &
BH-NS have neutrinos collide to gammas in clean fireball   \cr
\item & Meszaros et al.    & 1992 & MNRAS, 257, 29P       & NS & NS  & COS  &
NS-NS have neutrinos collide to gammas in clean fireball   \cr
\item & Cline et al.       & 1992 & ApJ, 401, L57         & BH &     & DISK &
Primordial BHs evaporating could account for short hard GRBs          \cr
\item & Rees et al.        & 1992 & MNRAS, 258, 41P       & NS & ISM & COS  &
Relativistic fireball reconverted to radiation when hits ISM          \cr
}

\bigskip\bigskip
\bigskip\bigskip

}

\noindent
Table from: Nemiroff, R. J. 1993, Comments on Astrophysics, 17, No. 4, in press

\vfill\eject
\end